\documentclass[aps,reprint]{revtex4-2}

\usepackage{graphicx}
\usepackage{dcolumn}
\usepackage{bm}

\begin{document}


\title{Soliton-mode proliferation induced by cross-phase modulation of harmonic waves by a dark-soliton crystal in optical media}


\author{E. Chenui Aban}
\affiliation{Laboratory of Research on Advanced Materials and Nonlinear Science (LaRAMaNS),
Department of Physics, Faculty of Science, University of Buea P.O. Box 63 Buea, Cameroon.}

\author{Alain M. Dikand\'e}
\email[Corresponding author: ]{dikande.alain@ubuea.cm}
\affiliation{Laboratory of Research on Advanced Materials and Nonlinear Science (LaRAMaNS),
Department of Physics, Faculty of Science, University of Buea P.O. Box 63 Buea, Cameroon.}

\date{\today}

\begin{abstract}
The generation of high-intensity optical fields from harmonic-wave photons, interacting via a cross-phase modulation with dark solitons both propagating in a Kerr nonlinear medium, is examined. The focus is on a pump consisting of time-entangled dark-soliton patterns, forming a periodic waveguide along the path of the harmonic-wave probe. It is shown that an increase of the strength of cross-phase modulation respective to the self-phase modulation, favors soliton-mode proliferation in the bound-state spectrum of the trapped harmonic-wave probe. The induced soliton modes, which display the structures of periodic soliton lattices, are not just rich in numbers, they also form a great diversity of population of soliton crystals with a high degree of degeneracy.
\end{abstract}


\maketitle

\section{\label{intro}Introduction}
Optical wave trapping, cloning, reconfiguration, duplication, parametric amplification and recompression \cite{1,2,3,4,5,6,7,8,9,9a,9b} are physical processes associated with wave interactions \cite{10,11,13} in nonlinear optical media. These processes are usually controlled by nonlinear phenomena originating from the intensity-dependent index of refraction of propagation media, which causes cross-phase or induced-phase modulations \cite{14,15} determining shape profiles of the propagating optical fields. Such processes find widespread applications in modern communication technology, and particularly in the processings of relatively low-power fields (such as harmonic fields) interacting with fields of sufficiently high intensity (such as optical solitons), leading to their cloning and reconfiguration into optical fields of higher powers \cite{9a}. \par
Since soliton cloning and reconfiguration can involve a sizable energy cost from the pump field, these processes have most often been envisaged between two solitons of slightly different powers. Thus in refs. \cite{5,16}, a reconfiguration scheme was proposed in which an intense pump beam with soliton features induces the focusing of a weaker probe beam of different wavelength, but also with soliton features, via cross-phase modulation. The underlying mechanism is simply understood by recalling that an optical soliton propagating in a Kerr nonlinear medium, creates a local distortion of the refractive index that travels with the soliton down the nonlinear propagation medium. As a result of this refractive index distortion a waveguide can be induced, that acts like a local potential by trapping and reshaping another much weaker pulse, different from the soliton pump in both frequency and polarization. However, in ref. \cite{9a}, Steiglitz and Rand suggested the possibility to use optical solitons propogating for instance in an optical fiber, to trap an reshape continuous-wave photons by means of their cross-phase modulation with the optical solitons. Thus, by considering a localized bright soliton pump interacting with an harmonic photon field, they established that the probe field was reshaped into new modes the eigenstates of which were described by a linear eigenvalue equation with a reflectionless potential. Subsequent to the study of Steiglitz and Rand \cite{9a}, the phenomenon of harmonic-wave trapping and reconfiguration by bright solitons was extended to the context of a waveguide created by a periodic train of bright solitons \cite{10} forming a bright-soliton crystal. This later study led to a linear eigenvalue problem of the Lam\'e type for the probe field \cite{17}, and its bound states were shown to form spectra of rich and abundant soliton modes. Namely in ref. \cite{9} it was established that increasing the strength of cross-phase modulation relative to the self-phase modulation, favors an increase of the population and the degeneracy of soliton modes composing bound-state spectra of the trapped probe. \par
While the dynamics of bright solitons as well as their stability under mutual collisions are relatively well understood, dark solitons have remained a curiosity for some reasons. Most importantly dark solitons are odd-symmetry structures, and for this reason they can only propagate in specific media \cite{18}. Nevertheless it is well established that dark solitons have simpler collision dynamics than their bright counterparts \cite{19}, are generally more stable against various perturbations \cite{20}, and hence may offer some important advantages in optical field processing applications. Based on this later features Steiglitz \cite{21} considered using a localized dark soliton pump to trap and reshape harmonic-wave probes. He found that probe modes induced by cross-phase modulation in the waveguide of the single dark soliton, were determined by a linear eigenvalue equation with a reflecting scattering potential. Because of this the groundstate was not a Goldstone translation of the pump as in the case involving a bright-soliton pump \cite{9a}, but instead a localized sech-type pulse soliton. \\
Motivated by results of a previous study \cite{9}, in which we found that a waveguide consisting of a crystal of bright solitons favors relatively more diverse and abundant soliton modes in the probe spectrum, in the present study we shall examine the problem of harmonic-wave reconfiguration by a waveguide consisting of a periodic train of dark solitons. Below we start with the presentation of the model, and obtain the periodic dark-soliton solution to the pump equation. With this solution we show that the probe equation can be formulated in terms of a Lam\'e-type eigenvalue problem, and derive some exact bounded states to this eigenvalue problem under specific conditions. 
\section{\label{sec2}The pump-probe equations and dark-soliton-crystal solution to the pump equation}
The propagation equations for the system composed of a nonlinear pump field, coupled to an harmonic-wave field via a cross-phase modulation anf propagating together in a Kerr nonlinear optical medium, are given by: 
\begin{eqnarray}
i\frac{\partial{v}}{\partial{z}} +\frac{\partial^{2}{v}}{\partial{t^2}} -2\zeta|v|^2 v&=&0, \label{eqa1} \\
i\frac{\partial{u}}{\partial{z}}+k_1\frac{\partial^{2}{u}}{\partial{t^2}}-2k_2|v|^2u&=&0. \label{eqa2}
\end{eqnarray}
In the first equation, which is precisely the cubic nonlinear Schr\"odinger equation with self-defocusing nonlinearity, the quantity $v$ is the pump envelope, $z$ is the propagation distance, $t$ is the propagation time and $\zeta$ is the coefficient of self-phase modulation. The quantity $u$ in the second equation is the probe envelope, $k_1$ is the coefficient of group-velocity-dispersion and $k_2$ is the coefficient of cross-phase modulation. \\
In eq. (\ref{eqa1}) the nonlinear coefficient is effectively negative i.e. $-\zeta$ with $\zeta>0$, corresponding to a Kerr optical medium with self-defocusing nonlinearity. In ref. \cite{21} the problem of harmonic-wave trapping and reshaping by a waveguide created by a single dark soliton, was considered. Here we are interested in the context when the waveguide is created by a period train of single dark solitons, forming a dark-soliton crystal. In this purpose we consider a solution to eq. (\ref{eqa1}) describing a stationary wave, $v(z,t)=A(t)exp[-i(kz-\omega t)]$, where $k$ is the wave number and $\omega$ is the frequency. Substituting this in eq. (\ref{eqa1}), we find that the wave amplitude $A(t)$ must obey the first-integral equation:
\begin{equation}
\frac{\partial{A}}{\partial{t}}=\zeta\left(\sqrt{A^4-\frac{s}{\zeta}A^2+\rho_1} \right), \label{Eq2}
\end{equation}
 with $\rho_1$ an energy contant determining shape profile of $A(t)$. Solving eq. (\ref{Eq2} with periodic boundary conditions \cite{22,23,24,25}, the pump amplitude $A(t)$ is found to be the following nonlocalized periodic pattern of time-entangled dark solitons:
\begin{equation}
A(t)=\frac{Q}{\sqrt{\zeta}}sn[Q(t-t_0)], \label{eq3}
\end{equation}
 where $sn()$ is a Jacobi elliptic function of modulus $\kappa$ (with $0\leq \kappa\leq 1$), and: 
\begin{equation}\label{eq4}
Q=\sqrt{\frac{s}{(1+\kappa^2)}}, \hskip 0.25truecm s=k-\omega^2.
\end{equation}\par
\begin{figure*} 
\begin{minipage}[c]{0.51\textwidth}
\includegraphics[width=3.in,height=2.in]{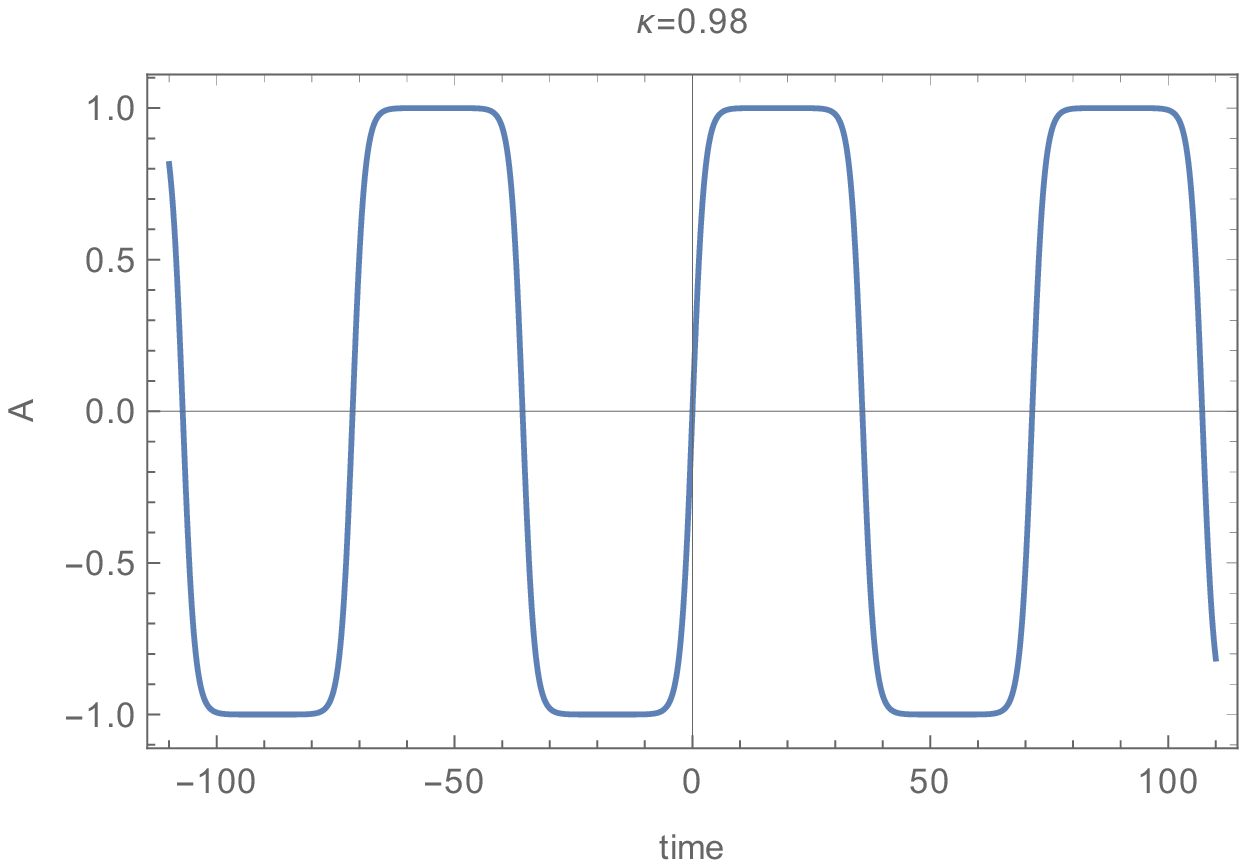}
\end{minipage}%
\begin{minipage}[c]{0.51\textwidth}
\includegraphics[width=3.in,height=2.in]{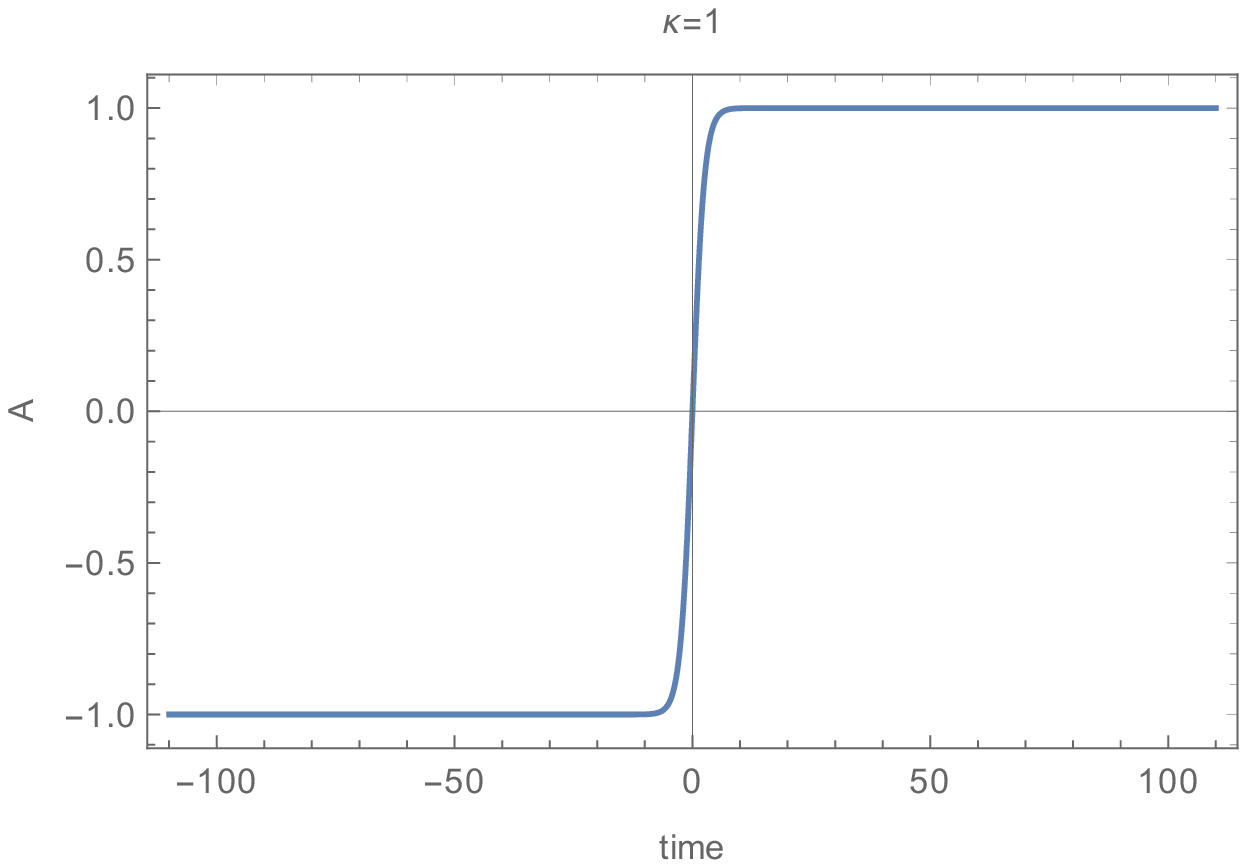}
\end{minipage} 
\caption{\label{Fig1}(Color online) Amplitude of the pump field $A(t)$ given by eq. (\ref{eq3}) versus time, for $\kappa=0.98$ (left graph) and $\kappa=1$ (right graph).} 
\end{figure*}

The amplitude $A(t)$ of the periodic dark soliton (\ref{eq3}) is represented in fig. \ref{Fig1}, for $\kappa=0.98$ (left graph) and $\kappa=1$ (right graph). Note that when $\kappa\rightarrow 1$ the Jacobi elliptic function $sn()\rightarrow tanh()$, corresponding to the dark soliton pump obtained in \cite{21}. 
\section{\label{sec3}Pump-induced trapping, reshaping and probe-mode proliferation}
Using the periodic dark soliton solution to the pump equation obtained in eq. (\ref{eq3}), we will now seek solutions to the probe equation (\ref{eqa2}). Proceeding with it is useful to start by the important remark that eq. (\ref{eqa2}) is a linear chr\"odinger equation, but with a time-dependent "external" potential represented by the norm squared of the pump envelope $q(z,t)$. Substituting (\ref{eq3}) in the probe equation given by (\ref{eqa2}, and expressing the probe envelope as a stationary wave i.e. $A(t)=u(t)exp(-iqz)$, where $u(t)$ is the core of the probe envelope and $q$ is its wave number, we obtain the Lam\'e equation \cite{17}:
\begin{equation}
\frac{\partial^{2}{u}}{\partial{\tau^2}}+\Bigg(P(q)-l(l+1)\kappa^2sn^2(\tau)\Bigg)u=0, \label{eq8}
\end{equation}
\begin{equation}
P(q)=\frac{q}{k_1Q^2}, \hskip 0.25truecm
\tau=Q(t-t_0), \hskip 0.25truecm
l(l+1)=\frac{2k_2}{\kappa^2k_1\zeta}. \label{eq9}
\end{equation}
It is worth stressing that when $\kappa=1$, the Lam\'e equation (\ref{eq9}) becomes the Associated Legendre equation obtained in ref. \cite{21}. \par The Lam\'e equation possesses a rich spectrum with a great variety of eigenmodes \cite{17}. However the most relevant to us are its eigenmodes that display a permanent profile typical of solitons. Precisely these later modes are bound states of the Lam\'e equation, and because their formation through the cross-phase modulation with the dark-soliton crystal involves energy cost (momentum transfer) from the pump, they can be looked out as low-energy states of the probe spectrum created by the pump-induced periodic potential (\ref{eq3}). Discrete states of Lam\'e's equation form a spectrum of finite orthogonal modes, whose population depends on the integer quantum number $l$ \cite{17}. According to equation (\ref{eq9}), values of the integer quantum number $l$ will be determined by the competition between the self-phase modulation responsible for the fiber nonlinearity, and the cross-phase modulation exerted by the pump field on the harmonic probe. For a given value of $l$, the discrete spectrum of Lam\'e equation possesses $2l+1$ modes some of which can be degenerate \cite{9,10,17}. \par
\begin{figure*} 
\begin{minipage}[c]{0.51\textwidth}
\includegraphics[width=3.in,height=2.in]{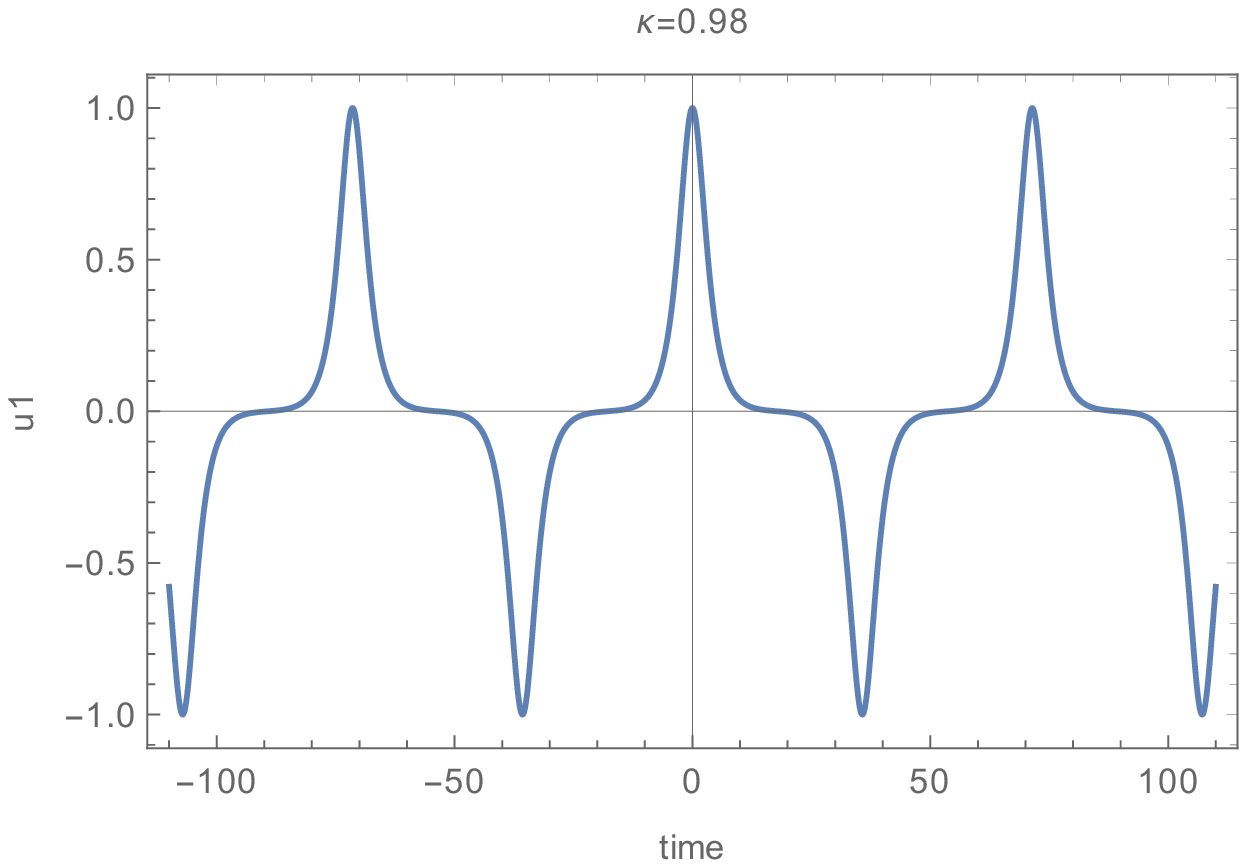}
\end{minipage}%
\begin{minipage}[c]{0.51\textwidth}
\includegraphics[width=3.in,height=2.in]{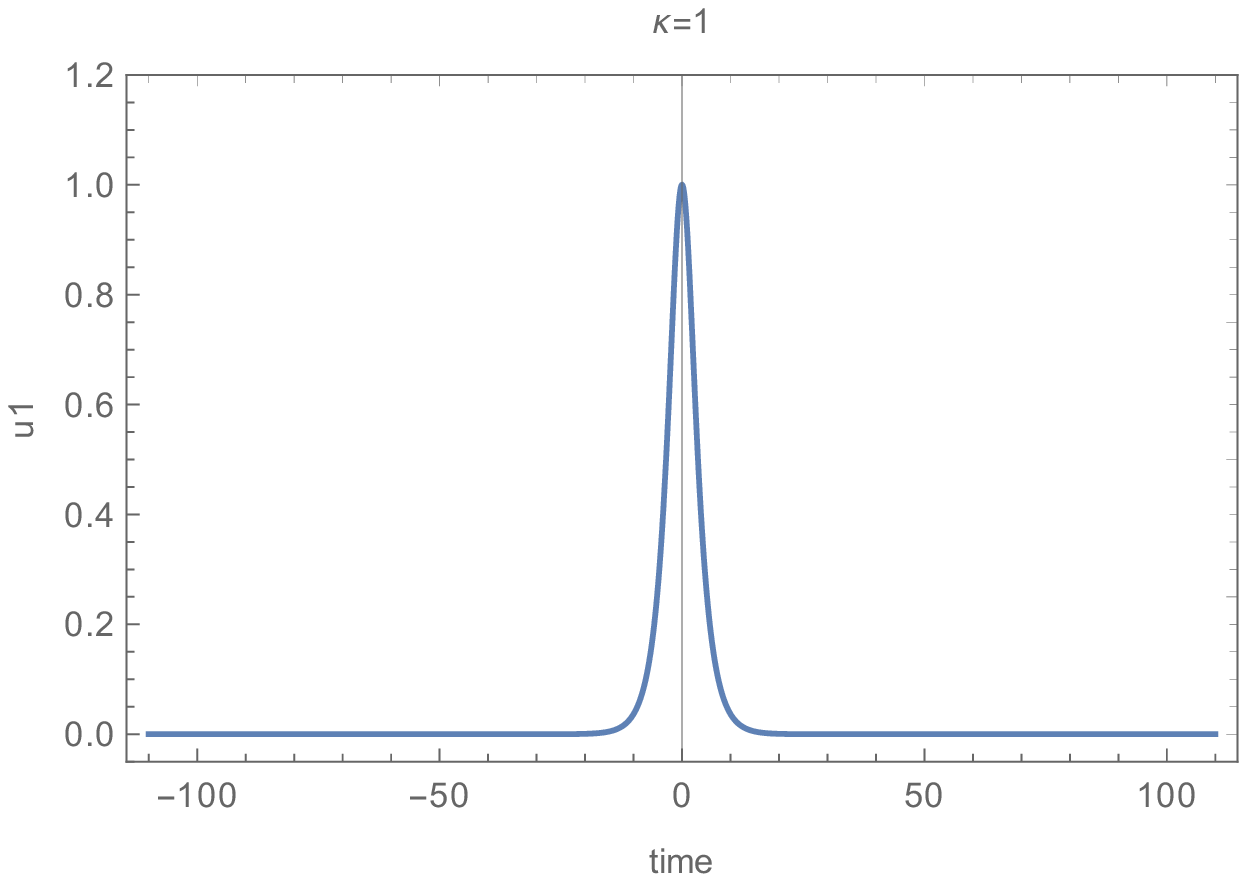}
\end{minipage}\vskip 0.25truecm
\begin{minipage}[c]{0.51\textwidth}
\includegraphics[width=3.in,height=2.in]{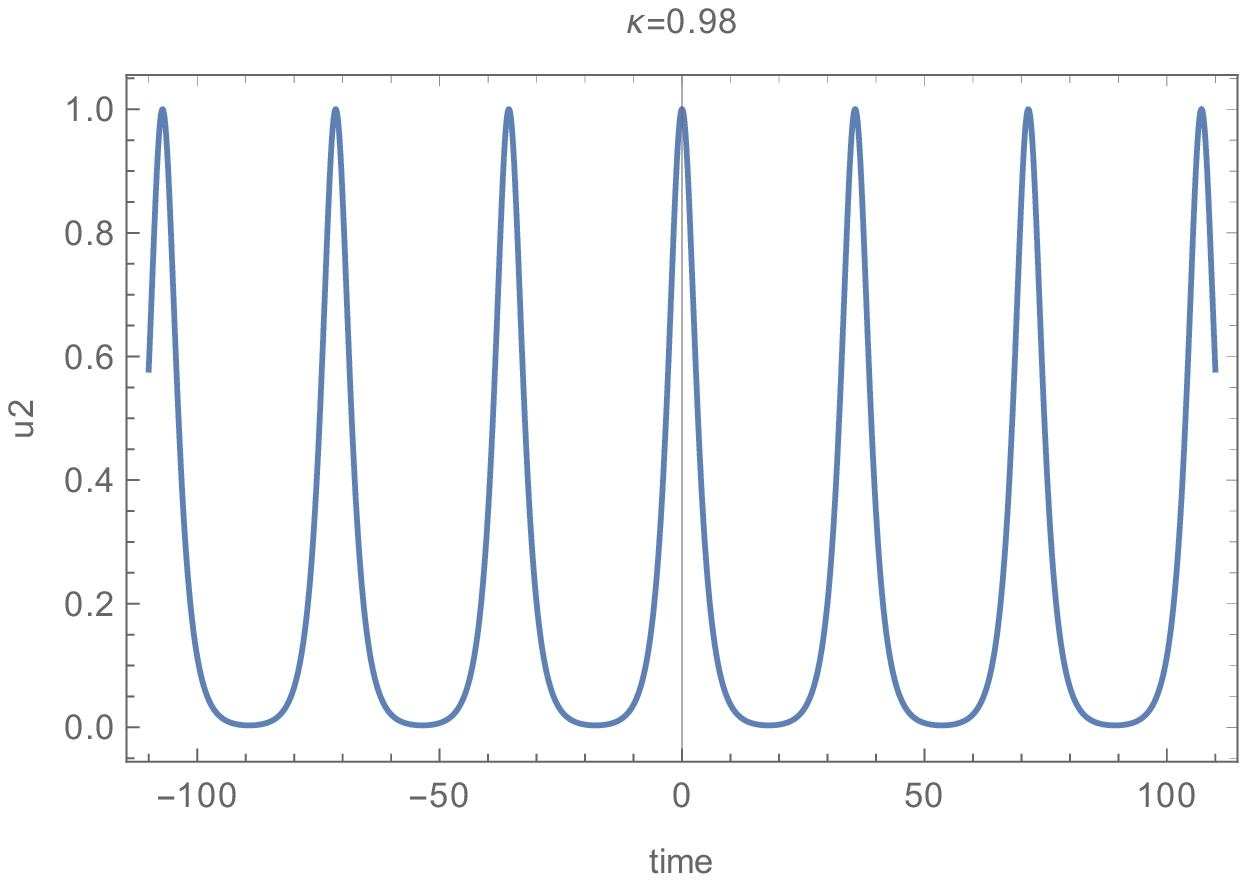}
\end{minipage}%
\begin{minipage}[c]{0.51\textwidth}
\includegraphics[width=3.in,height=2.in]{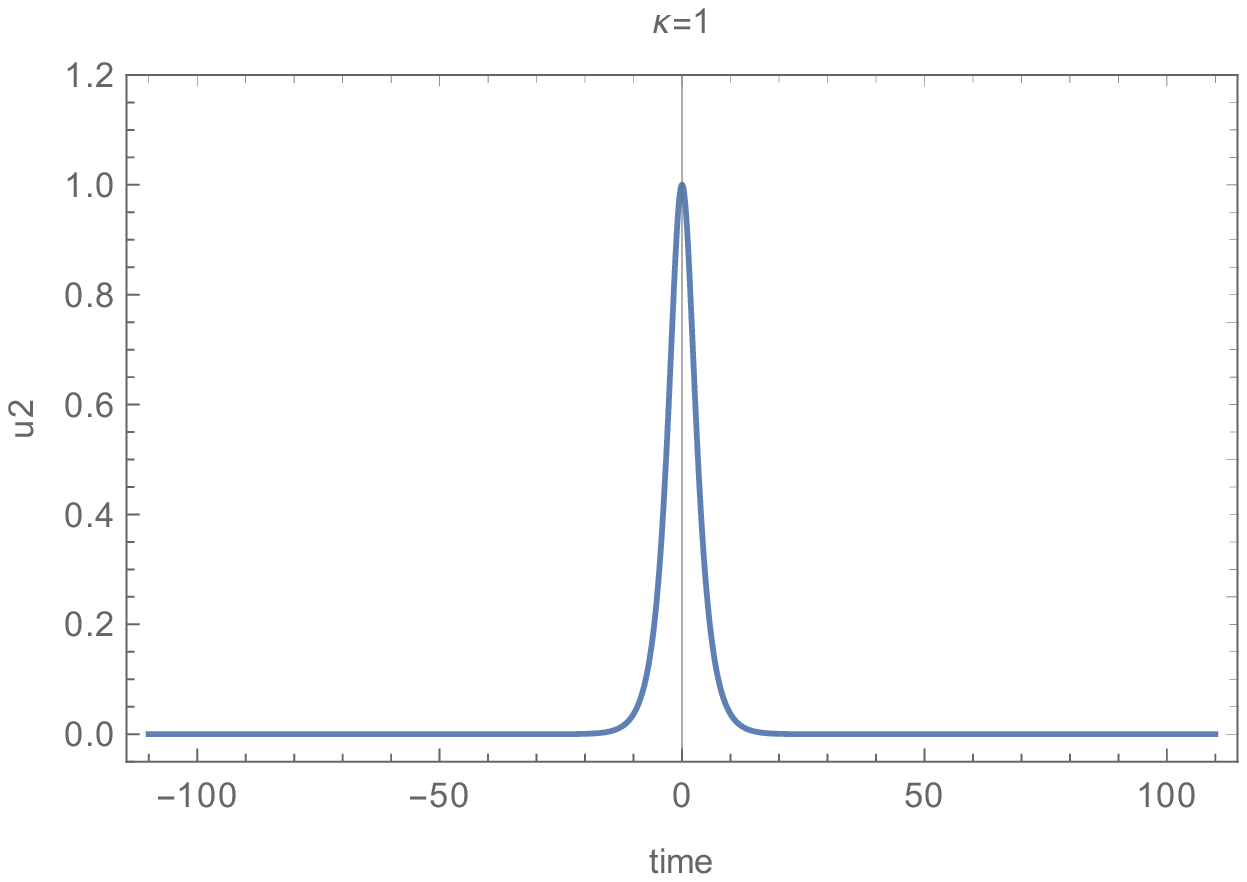}
\end{minipage}\vskip 0.25truecm
\begin{minipage}[c]{0.51\textwidth}
\includegraphics[width=3.in,height=2.in]{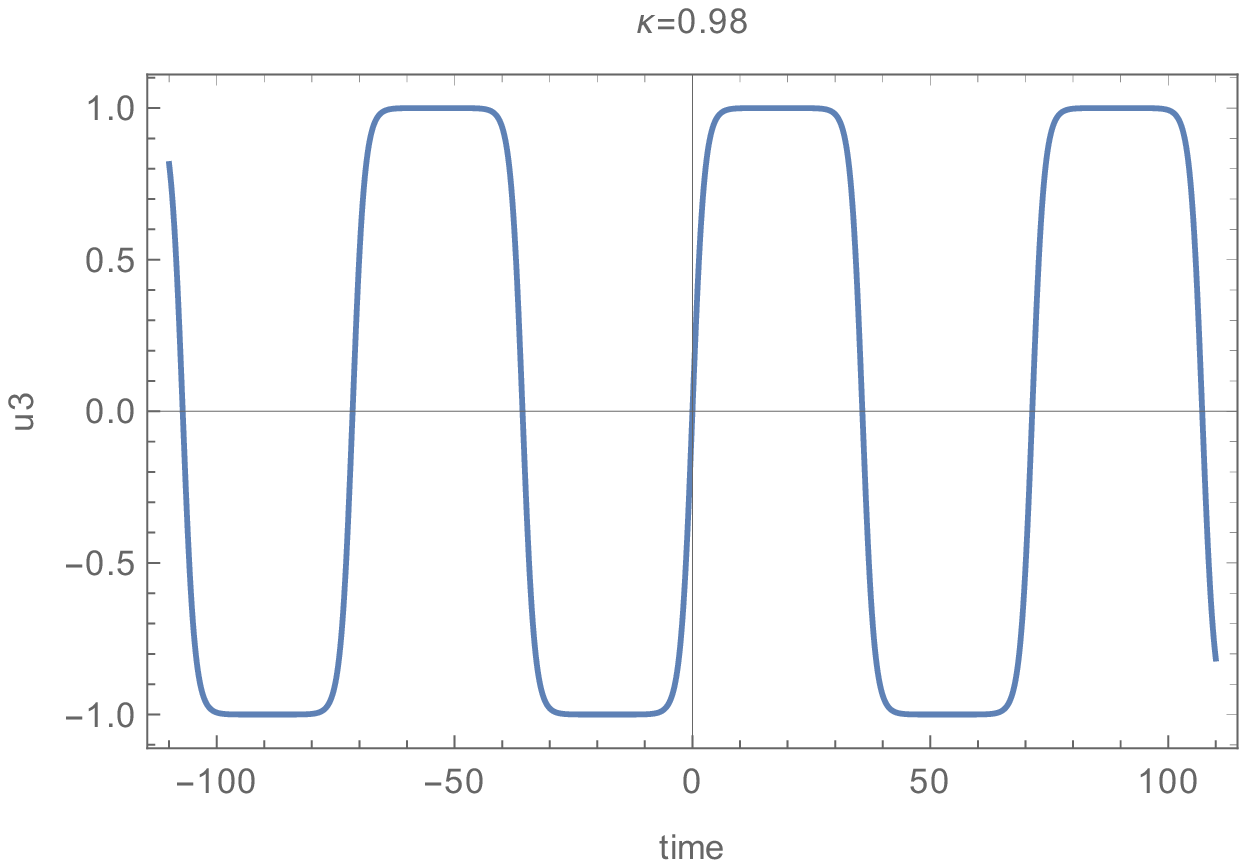}
\end{minipage}%
\begin{minipage}[c]{0.51\textwidth}
\includegraphics[width=3.in,height=2.in]{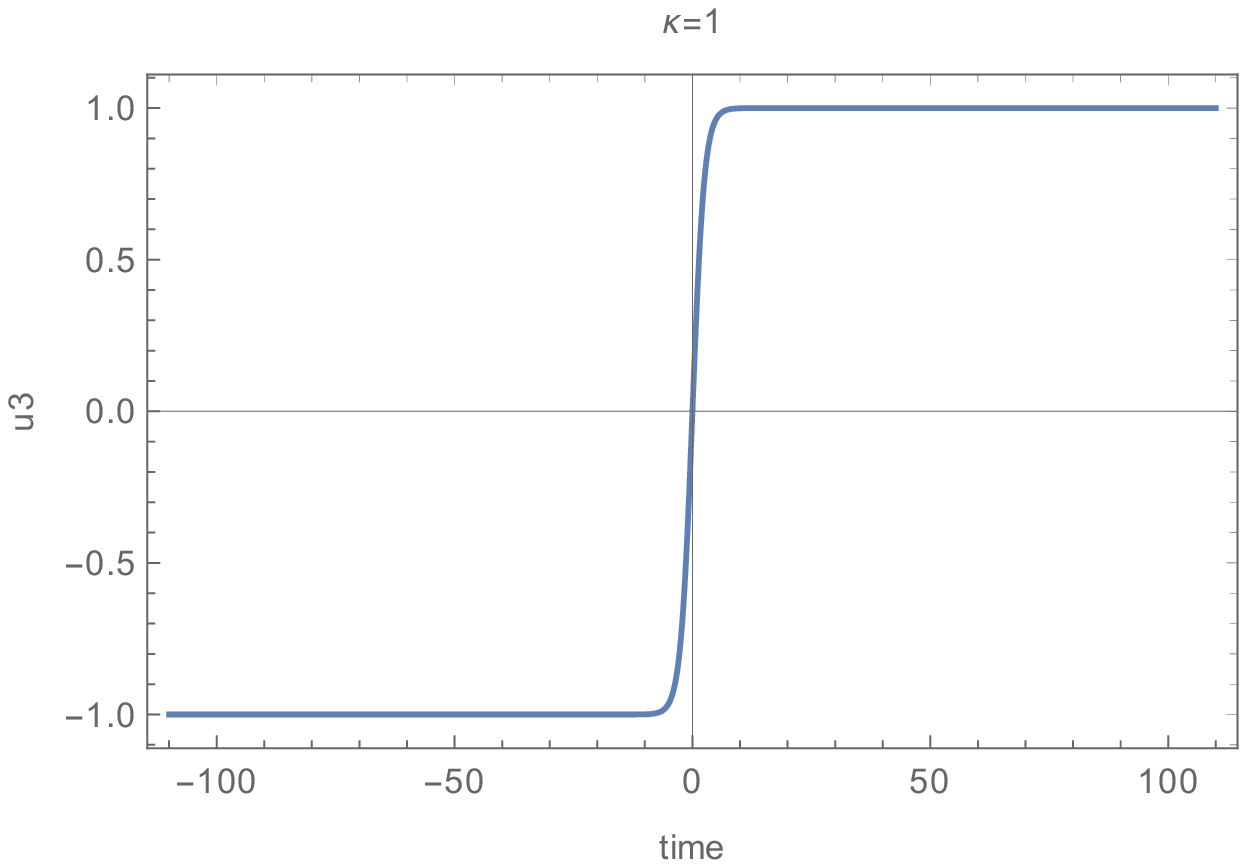}
\end{minipage} 
\caption{\label{Fig2}(Color online) Amplitudes of the three probe modes corresponding to $l=1$ versus time, for $\kappa=0.98$ (left graphs) and $\kappa=1$ (right graphs). $u_i$ in the graphs mean $u_{1i}$ in eqs. (\ref{eq12})-(\ref{eq14}), with $i=1,2,3$.} 
\end{figure*}

We start with the lowest value of $l$; $l=1$ corresponding to the case when the cross-phase modulation and the self-phase modulation coefficients are related by $k_2=k_1\zeta\kappa^2$. In this case the Lam\'e equation possesses three distinct localized modes, namely:
\begin{eqnarray}
u_{11}(\tau)&=&u^{(11)}cn(\tau),  \hskip 0.25truecm q=q_{11}=\frac{k_2Q^2}{\kappa^2\zeta}, \label{eq12} \\
u_{12}(\tau)&=&u^{(12)}dn(\tau),  \hskip 0.25truecm q=q_{12}=\frac{k_2Q^2}{\zeta}, \label{eq13} \\
u_{13}(\tau)&=&u^{(13)}sn(\tau),  \hskip 0.25truecm q=q_{13}=\frac{(1+\kappa^2)k_2Q^2}{\kappa^2\zeta}, \label{eq14} 
\end{eqnarray}
where $u^{(1i)}$ are normalization constants. The three modes are represented in fig.\ref{Fig2}, for $\kappa=0.98$ (left column) and $\kappa=1$ (right column).

\begin{figure*}
\begin{center}
\includegraphics[width=15.0cm,height=16.0 cm]{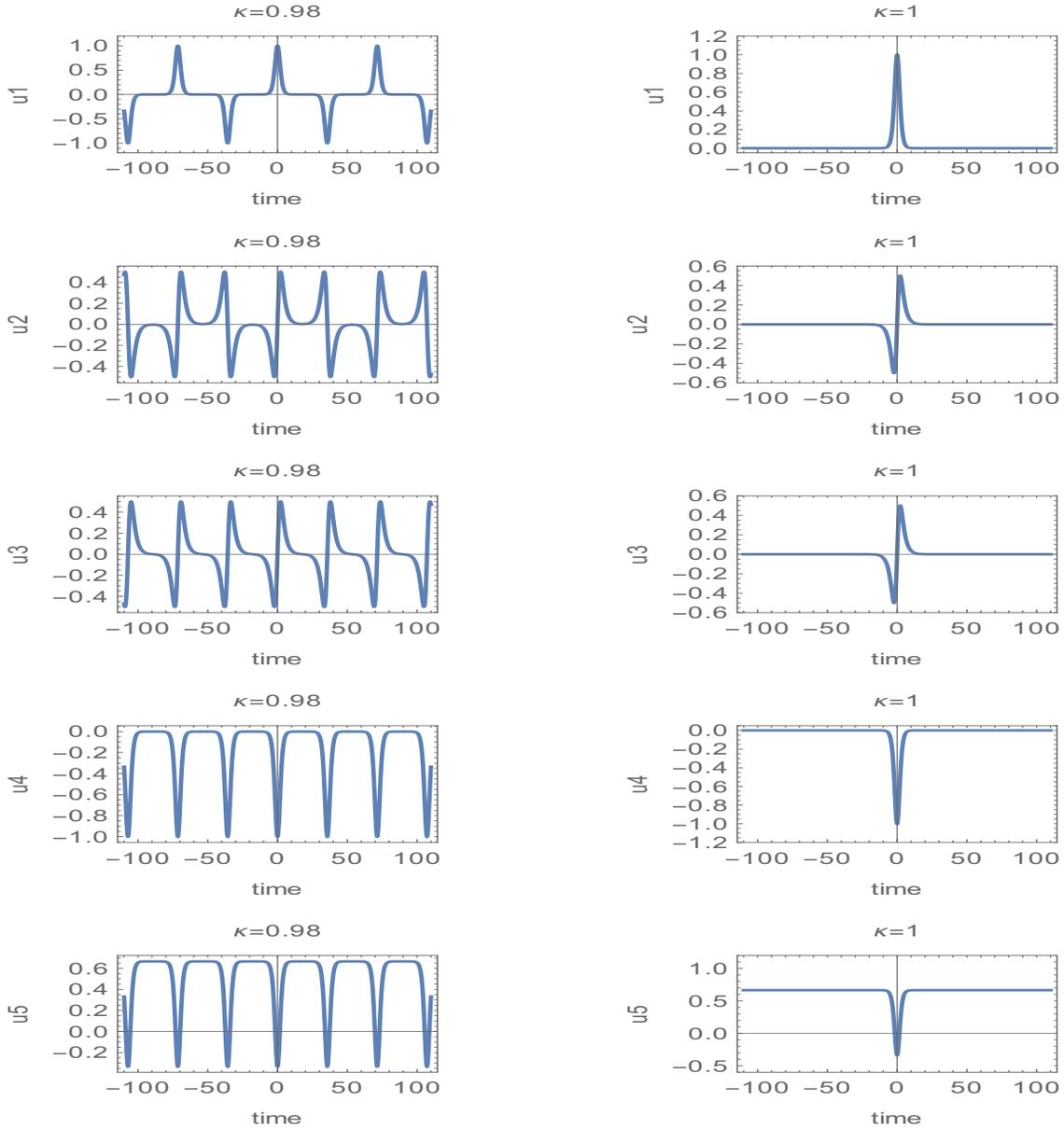}
\end{center}
\caption{\label{Fig3}(Color online) Temporal profiles of amplitudes of the five bounded modes given by eqs. (\ref{eq15})-(\ref{eq19}), for $\kappa=0.98$ (left column) and $\kappa=1$ (right column). $u_i$ in the graphs correspond to $u_{1i}$ in eqs. (\ref{eq15})-(\ref{eq19}), with $i=1,2,3,4,5$.}
\end{figure*}

 Taking $l=2$, or equivalently $k_2=3k_1\zeta\kappa^2$, leads to five distinct localized modes for the probe which are listed below:
\begin{equation}
u_{21}(\tau)=u^{(21)}cn(\tau)dn(\tau),  \hskip 0.2truecm q=q_{21}=\left(\frac{1+\kappa^2}{3\kappa^2}\right)\frac{k_2Q^2}{\zeta}, \label{eq15} 
\end{equation}
\begin{equation}
u_{22}(\tau)=u^{(22)}sn(\tau)dn(\tau),  \hskip 0.2truecm q=q_{22}=\left(\frac{1+4\kappa^2}{3\kappa^2}\right)\frac{k_2Q^2}{\zeta}, \label{eq16} 
\end{equation}
\begin{equation}
u_{23}(\tau)=u^{(23)}sn(\tau)cn(\tau),  \hskip 0.2truecm q=q_{23}=\left(\frac{4+\kappa^2}{3\kappa^2}\right)\frac{k_2Q^2}{\zeta}, \label{eq17} 
\end{equation}
\begin{eqnarray}
u_{24}(\tau)&=&u^{(24)}\left[sn^2(\tau)-\frac{1+\kappa^2+\sqrt{1-\kappa^2(1-\kappa^2)}}{3\kappa^2}\right], \nonumber \\ 
q=q_{24}&=&\left[\frac{2(1+\kappa^2)-\sqrt{1-\kappa^2(1-\kappa^2)}}{3\kappa^2}\right]\frac{k_2Q^2}{\zeta}, \label{eq18} 
\end{eqnarray}
\begin{eqnarray}
u_{25}(\tau)&=&u^{(25)}\left[sn^2(\tau)-\frac{1+\kappa^2-\sqrt{1-\kappa^2(1-\kappa^2)}}{3\kappa^2}\right],  \nonumber \\
q=q_{25}&=&\left[\frac{2(1+\kappa^2)+\sqrt{1-\kappa^2(1-\kappa^2)}}{3\kappa^2}\right]\frac{k_2Q^2}{\zeta}. \label{eq19}
\end{eqnarray}
The five bounded modes are plotted versus time in fig.\ref{Fig3}, for $\kappa=0.98$ (left column) and $\kappa=1$ (right column). 
 \par The third and last case considered is $l=3$, corresponding to $k_2=6k_1\zeta\kappa^2$. In this case the bound-state spectrum of the probe comprises seven distinct localized modes i.e.:\par
\begin{eqnarray}
u_{31}(\tau)&=&u^{(31)}sn(\tau)cn(\tau)dn(\tau), \nonumber \\ q=q_{31}&=&\left(\frac{2+2\kappa^2)}{3\kappa^2}\right)\frac{k_2Q^2}{\zeta}, \label{eq20}
\end{eqnarray}
\begin{eqnarray}
u_{32}(\tau)&=&u^{(32)}\left[sn^3(\tau)-\frac{2(1+\kappa^2)-\sqrt{4-7\kappa^2+4\kappa^2}}{5\kappa^2}sn(\tau)\right], \nonumber \\ 
q=q_{32}&=&\left[\frac{5(1+\kappa^2)+2\sqrt{4-7\kappa^2+4\kappa^2}}{6\kappa^2}\right]\frac{k_2Q^2}{\zeta}, \label{eq21}
\end{eqnarray}
\begin{eqnarray}
u_{33}(\tau)&=&u^{(33)}\left[sn^3(\tau)-\frac{2(1+\kappa^2)+\sqrt{4-7\kappa^2+4\kappa^2}}{5\kappa^2}sn(\tau)\right], \nonumber \\ 
q=q_{33}&=&\left[\frac{5(1+\kappa^2)-2\sqrt{4-7\kappa^2+4\kappa^2}}{6\kappa^2}\right]\frac{k_2Q^2}{\zeta}, \label{eq22}
\end{eqnarray}
\begin{eqnarray}
u_{34}(\tau)&=&u^{(34)}\,cn(\tau)\left[sn^2(\tau)-\frac{2+\kappa^2-\sqrt{4-\kappa^2(1-\kappa^2)}}{5\kappa^2}\right], \nonumber \\  
q=q_{34}&=&\left[\frac{5+2\kappa^2+2\sqrt{4-\kappa^2(1-\kappa^2)}}{6\kappa^2}\right]\frac{k_2Q^2}{\zeta}, \label{eq23}
\end{eqnarray}
\begin{eqnarray}
u_{35}(\tau)&=&u^{(35)}\,cn(\tau)\left[sn^2(\tau)-\frac{2+\kappa^2+\sqrt{4-\kappa^2(1-\kappa^2)}}{5\kappa^2}\right], \nonumber \\  
q=q_{35}&=&\left[\frac{5+2\kappa^2-2\sqrt{4-\kappa^2(1-\kappa^2)}}{6\kappa^2}\right]\frac{k_2Q^2}{\zeta}, \label{eq24}
\end{eqnarray}
\begin{eqnarray}
u_{36}(\tau)&=&u^{(36)}\,dn(\tau)\left[sn^2(\tau)-\frac{1+\kappa^2-\sqrt{1-\kappa^2+4\kappa^4}}{5\kappa^2}\right], \nonumber \\
q=q_{36}&=&\left[\frac{2+5\kappa^2+2\sqrt{1-\kappa^2+4\kappa^4}}{6\kappa^2}\right]\frac{k_2Q^2}{\zeta}, \label{eq25}
\end{eqnarray}
\begin{eqnarray}
u_{37}(\tau)&=&u^{(37)}\,dn(\tau)\left[sn^2(\tau)-\frac{1+\kappa^2+\sqrt{1-\kappa^2+4\kappa^4}}{5\kappa^2}\right], \nonumber \\
q=q_{37}&=&\left[\frac{2+5\kappa^2-2\sqrt{1-\kappa^2+4\kappa^4}}{6\kappa^2}\right]\frac{k_2Q^2}{\zeta}. \label{eq26}
\end{eqnarray}
Although the analytical expressions of the seven modes seem to suggest complex combinations of Jacobi elliptic functions, we can convince ourselves of the contrary by examining their expressions for $\kappa=1$, a value for which these analytical expressions are fundamental components composing the soliton trains in the seven distinct modes. This remark also holds for the cases $l=1$ and $l=2$. To gain a better understanding of shape profiles of these fundamental components, in tables \ref{tab1}, \ref{tab2} and \ref{tab3} their eigenfunctions $u(\tau)$ are listed together with corresponding eigenvalues $q$.

\begin{table}
\caption{\label{tab1}Fundamental components (solutions with $\kappa=1$) of the $l=1$ eigenmodes.}
\begin{ruledtabular}
\begin{tabular}{cc}
Eigenfunction & Eigenvalue\\
\hline
$u_{11}(\tau)=u_{12}(\tau)\propto sech(\tau)$ & $q_{11}=q_{12}=k_2Q^2/\zeta$ \\
$u_{13}(\tau)\propto \tanh(\tau)$ & $q_{13}=2k_2Q^2/\zeta$ \\
\end{tabular}
\end{ruledtabular}
\end{table}

\begin{table}
\caption{\label{tab2}Fundamental components (solutions with $\kappa=1$) of the $l=2$ eigenmodes.}
\begin{ruledtabular}
\begin{tabular}{cc}
Eigenfunction & Eigenvalue\\
\hline
$u_{21}(\tau)\propto sech^2(\tau)$ &  $q_{21}=2k_2Q^2/(3\zeta)$ \\
$u_{22}(\tau)=u_{23}(\tau)\propto sech(\tau)\tanh(\tau)$ & $q_{22}=q_{23}=5k_2Q^2/(3\zeta)$ \\
$u_{24}(\tau)\propto -sech^2(\tau)$ & $q_{24}=k_2Q^2/\zeta$ \\
$u_{25}(\tau)\propto 2-3sech^2(\tau)$ & $q_{25}=5k_2Q^2/(3\zeta)$ \\
\end{tabular}
\end{ruledtabular}
\end{table}

\begin{table}
\caption{\label{tab3}Fundamental components (solutions with $\kappa=1$) of the $l=3$ eigenmodes.}
\begin{ruledtabular}
\begin{tabular}{cc}
Eigenfunction & Eigenvalue\\
\hline
$u_{31}(\tau)\propto sech^2(\tau)\tanh(\tau)$ &  $q_{31}=4k_2Q^2/(3\zeta)$ \\
$u_{32}(\tau)\propto \Big(2-5sech^2(\tau)\Big)\tanh(\tau)$ &  $q_{32}=2k_2Q^2/\zeta$ \\
$u_{33}(\tau)\propto -sech^2(\tau)\tanh(\tau)$ &  $q_{33}=4k_2Q^2/(3\zeta)$ \\
$u_{34}(\tau)\propto \Big(4-5sech^2(\tau)\Big)sech(\tau)$ &  $q_{34}=11k_2Q^2/(6\zeta)$ \\
$u_{35}(\tau)\propto -sech^3(\tau)$ &  $q_{35}=k_2Q^2/(2\zeta)$ \\
$u_{36}(\tau)\propto sech(\tau)\tanh^2(\tau)$ &  $q_{36}=11k_2Q^2/(6\zeta)$ \\
$u_{37}(\tau)\propto \Big(1-5sech^2(\tau)\Big)sech(\tau)$ &  $q_{37}=k_2Q^2/(2\zeta)$ \\
\end{tabular}
\end{ruledtabular}
\end{table}

As it is apparently, the three tables feature very rich and varied spectra of bounded states for the three values of $l$ considered. Also remarkable, table \ref{tab1} suggests that the $l=1$ spectrum possesses two modes which are nearly degenerate, whereas the $l=2$ spectrum has three nearly degenerate modes (see table \ref{tab2}) and the spectrum for $l=3$ possesses three distinct double-degenerate modes as one can see in table \ref{tab3}, where only $u_{32}$ out of the seven bounded states is non degenerate.   
\section{\label{sec4}Conclusion}
 We examined profiles of an harmonic wave propagating in the waveguide structure created by a periodic lattice of single dark solitons, obtained as a periodic solution to the self-defocusing cubic nonlinear Schr\"odinger equation describing the propagation of a pump field along an optical fibre. In ref. \cite{21} the same problem was discussed assuming that the harmonic probe is coupled to a single dark soliton. We obtained that due to the coupling of the pump and probe via the cross-phase modulation, the probe equation can be transformed into a general family of eigenvalue equation called Lam\'e equation. Considering bounded states of this eigenvalue problem, we found that the population of the associated discrete spectrum was determined by an integer quantum number which in turn was determined by the competition between the self-phase moduation and the cross-phase modulation. We obtained analytical expressions of these bounded modes of the trapped probe for $l=1,2,3$, and observed that as $l$ increases the number of bounded modes was higher and higher and the spectra more and more degenerate. 
\par Our study demonstrates unambiguously that waveguides induced by a periodic train of dark solitons, can be used to control weak probes in the same way bright solitons can. Dark solitons offer real advantages over bright soliton collisions in controlling light waves: Firstly dark solitons are well known to be more stable in the presence of noise \cite{18,19,20}, and are generally more robust than bright solitons. Secondly the probe, which is of much lower intensity and here an harmonic wave, is expected to peak at the dip in the intensities of its host single-dark soliton components, thus increasing the signal-to-noise ratio and making it easier, in principle, to detect. The present study finds its most important application in quantum communication and cryptographic systems, we think in particular of experimental verification of photon capture and transport in an optical fiber. The problem of detecting the probe in the presence of a pump, and tailoring physical parameters to realize virtual devices, is also of current interest in quantum computing applications involving entangled photon emitters and photon qubits \cite{m1,m2,m3,m4}.

\begin{acknowledgments}
A. M. Dikand\'e thanks the Alexander von Humboldt foundation for support, and the Max-Planck Institute for the Physics of Complex Systems (MPIPKS), Dresden where part of this work was done.
\end{acknowledgments}

\end{document}